\documentclass[aps,prl,10pt,twocolumn,superscriptaddress,nofootinbib,floatfix]{revtex4-2}
\usepackage[utf8]{inputenc}
\usepackage[T1]{fontenc}
\usepackage{amsmath,amssymb}
\usepackage{graphicx}
\usepackage[dvipsnames]{xcolor}
\usepackage{booktabs}
\usepackage[colorlinks=true,linkcolor=blue!60!black,citecolor=blue!60!black,urlcolor=blue!60!black]{hyperref}
\newcommand{\dd}{\mathrm{d}}
\newcommand{\Li}{\mathrm{Li}}
\newcommand{\anc}{\textup{[anc,\,\citenum{Ancillary}]}}
\newif\ifblue \bluefalse
\ifblue
\hypersetup{allcolors=black}
\makeatletter
\newcommand{\new}[1]{{\color{blue}\def\@citecolor{blue}\def\@linkcolor{blue}#1}}
\makeatother

\newcommand{\bluehead}{\color{blue}}
\else
\newcommand{\new}[1]{#1}

\newcommand{\bluehead}{}
\fi
\newcommand{\sechead}[1]{\textit{#1}}
\newcommand{\EM}[1]{\new{\hyperlink{em:#1}{[EM-#1]}}}
\newcommand{\emhead}[2]{\par\noindent\setcounter{equation}{0}%
  \renewcommand{\theequation}{#1\arabic{equation}}\renewcommand{\theHequation}{EM#1.\arabic{equation}}%
  \hypertarget{em:#1}{}\sechead{\new{#1.} #2}---\ignorespaces}

\begin{document}

\title{Massive Triangle in Cosmological Correlator beyond Conformal Coupling}

\author{Taushif Ahmed}
\email{taushif.ahmed@ur.de}
\affiliation{Institute for Theoretical Physics, University of Regensburg, 93040 Regensburg, Germany}
\author{Syed Mehedi Hasan}
\email{syed-mehedi.hasan@ur.de}
\affiliation{Institute for Theoretical Physics, University of Regensburg, 93040 Regensburg, Germany}

\begin{abstract}
A massive scalar coupled in pairs to the inflaton generates a one-loop primordial bispectrum whose mass corrections at generic momenta have lacked closed analytic expressions beyond conformal coupling. We obtain two leading mass corrections at arbitrary external momenta in terms of multiple polylogarithms of uniform weights three and four. Auxiliary-energy dressings preserve rational time integrals, revealing unexpected simplicity despite elliptic intermediate geometry. Independent full-loop evaluations, including blinded predictions, validate the closed forms. We also determine a subleading light-field singularity and recover the heavy-mass effective theory.
\end{abstract}

\maketitle

\textit{Introduction}---A massive field during inflation can leave its imprint on primordial correlations, offering a cosmological probe of particle physics at the Hubble scale $H$ \cite{Chen:2009zp,Baumann:2011nk,Noumi:2012vr,Arkani-Hamed:2015bza,Lee:2016vti,Arkani-Hamed:2018kmz}. Primordial non-Gaussianity is already constrained by observations \cite{Planck:2019kim}. Yet some particles are invisible to tree-level probes: a $\mathbb Z_2$-odd scalar $\sigma$, or a charged field, couples to the inflaton only in pairs. Loop corrections in inflation have a long history \cite{Weinberg:2005vy,Senatore:2009cf}; for such paired couplings, the leading bispectrum includes a finite triangle with three massive propagators \cite{Chen:2016uwp,Chen:2016hrz,Kumar:2017ecc,Qin:2026yuo,Borinsky:2026roj}. Unlike an ordinary scattering amplitude, this triangle involves several time orderings, nonconserved vertex energies, and Bessel-function propagators. Its analytic structure has remained a central obstacle to computing massive cosmological loops.

Analytic progress has followed complementary routes. Massive-field bubbles have been obtained analytically \cite{Xianyu:2022jwk,Liu:2024xyi,Liu:2026jzn}, and a boundary triangle can arise from a field redefinition \cite{Wang:2025qfh}; spectral and dispersive methods offer further access to massive loops \cite{Xianyu:2022jwk,Liu:2024xyi,Liu:2026jzn}. At conformal coupling, $m^2=2H^2$, the rescaled scalar modes obey the massless flat-space equation---the unique such mass in the minimally coupled theory. The propagators become elementary, permitting analytic one-loop results \cite{Heckelbacher:2022hbq,Lee:2023jby,Chowdhury:2023arc,Chowdhury:2025dress,Bhowmick:2025sing,Pimentel:2026kqc,Su:2026jhc}, including a compact weight-two polylogarithmic formula for the triangle \cite{Henn:2026lfz}. Its wavefunction master integrals nevertheless contain elliptic sectors \cite{Benincasa:2024ptf}, whose periods cancel in the corresponding flat-space wavefunction coefficient \cite{McCune:2026qfy}. This illustrates the unexpected simplicity of cosmological correlators \cite{ArkaniHamed:2026simpler,Chowdhury:2026simp}.

Those closed conformal triangle calculations involve non-derivative scalar interactions without inflaton legs. For the kinetic portal, however, a differentiated inflaton leg supplies the same conformal-time power as two conformally coupled scalar legs; consequently, the time integrals agree at $m^2=2H^2$ after identifying each vertex energy with the external inflaton momentum. Away from conformal coupling, the internal propagators involve Bessel functions of generic order, and the time integrals no longer reduce to rational functions of the energies. The physical triangle has been evaluated numerically at two heavy masses \cite{Borinsky:2026roj}. Its leading soft signal \cite{Qin:2026yuo} and leading light-field singularity \cite{An:2018tcq} were also known. McCune obtained a finite Mellin--Barnes representation for the complete triangle at general masses and a convergent squeezed expansion for its collider signal \cite{McCune:2026vqb}. Ref.~\cite{McCune:2026oneloop} gives the exact squeezed coefficients and corrections away from the squeezed limit. The complete correlator, however, has not been expressed in closed polylogarithmic form beyond conformal coupling. Differential equations organize tree-level cosmological correlators \cite{Arkani-Hamed:2023kig,Arkani-Hamed:2023bsv}, where mass corrections can increase polylogarithmic weight \cite{Gasparotto:2024bku,Baumann:2026atn,Ahmed:2026triple}; the analytic function space of the complete massive triangle therefore remained unknown.

\emph{We now determine the first two such corrections in closed form, for arbitrary external momenta.} The first is a weight-three polylogarithm with maximal-cut prefactors; the second is a weight-four multiple polylogarithm with rational prefactors controlled by residual cuts. Surprisingly, both retain the conformal alphabet despite elliptic curves in intermediate integrations. We derive the coefficients through energy dressings of the massive propagators, validate the results against independent complete-loop integrations and exact folded limits, and connect the light- and heavy-mass regimes. Thus a genuinely massive cosmological loop exhibits, through second order, the uniform-weight structure of canonical Feynman integrals \cite{Henn:2013pwa}.

\begin{figure}[t]
\centering
\includegraphics[width=\columnwidth]{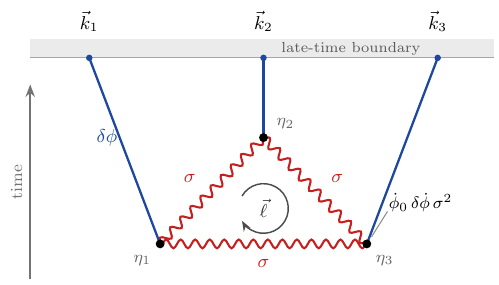}
\caption{The one-loop triangle. Each inflaton leg (blue)\new{, of momentum $\vec k_v$,} runs from the late-time boundary to a vertex
$\dot\phi_0\,\delta\dot\phi\,\sigma^2$ at its own conformal time $\eta_v$, and the massive field $\sigma$ (red)
closes the loop with momentum $\vec\ell$.}
\label{fig:triangle}
\end{figure}

\textit{\bluehead Setup and result}---We consider a scalar $\sigma$ of mass $m$, odd under a $\mathbb Z_2$ symmetry
$\sigma\to-\sigma$, coupled to the inflaton $\phi$ through the kinetic portal
\cite{Qin:2026yuo,Borinsky:2026roj,McCune:2026vqb},
\begin{equation}
\frac{\mathcal L}{\sqrt{-g}}\supset-\frac12(\partial\sigma)^2-\frac12m^2\sigma^2
+\frac{c}{\Lambda^2}(\partial\phi)^2\sigma^2\,,
\label{eq:L}
\end{equation}
with metric signature $(-,+,+,+)$. Since $\sigma$ appears only in pairs, it cannot mix with the inflaton, and its
leading signal arises at one loop. Writing $\phi=\phi_0(t)+\delta\phi$, the portal contains
$(\partial\phi)^2=-\dot\phi_0^2-2\dot\phi_0\,\delta\dot\phi+(\partial\delta\phi)^2$. The first term shifts the mass
of $\sigma$ and is absorbed into $m$; the second gives the cubic vertex $\dot\phi_0\,\delta\dot\phi\,\sigma^2$ at each
corner of the triangle (Fig.~\ref{fig:triangle}); the third gives the quartic vertex of the bubble. The inflaton
fluctuation is related to the Goldstone boson $\pi$ of broken time translations by $\delta\phi=\dot\phi_0\,\pi$ at
linear order \cite{Cheung:2007st}.

With $\tilde\phi=\delta\phi/H$, the triangle of Fig.~\ref{fig:triangle} contributes
\begin{equation}
\langle\tilde\phi_{\vec k_1}\tilde\phi_{\vec k_2}\tilde\phi_{\vec k_3}\rangle'_{\rm tri}
=\Big(\frac{c\,|\dot\phi_0|}{\Lambda^2}\Big)^3\,\frac{-s_\phi\,\mathcal C(k_1,k_2,k_3;\nu)}{\pi^2k_1k_2k_3}\,,
\label{eq:def}
\end{equation}
in the conventions of Ref.~\cite{Borinsky:2026roj}, with $s_\phi=-\dot\phi_0/|\dot\phi_0|$. The function $\mathcal C$
is homogeneous of degree $-3$ in the momenta. With $\nu=\sqrt{9/4-m^2/H^2}$, the conformally coupled mass is $\nu=\tfrac12$ and the
massless point $\nu=\tfrac32$. At $\nu=\tfrac12$ each internal line reduces to a flat-space line \cite{Arkani-Hamed:2017fdk}, and the composite
operator $\sigma^2$ has dimension two, so the triangle couples the inflaton to a conformal sector
\cite{Green:2013rd}. With $e_1=k_1+k_2+k_3$ we use $h_i=e_1-2k_i$, which vanishes at the folded
configuration $k_i=k_j+k_l$, and the ratios $x_i=2k_i/e_1$ and $y_i=h_i/e_1=1-x_i$.

At conformal coupling, the triangle is a known weight-two function:
its massless flat-space in--in representation was evaluated in
Ref.~\cite{Pimentel:2026kqc}, and the compact de Sitter form was derived in
Ref.~\cite{Henn:2026lfz}. Specializing the latter to one inflaton leg per
vertex, $X_i=k_i$, gives in our normalization
\begin{align}
\mathcal C_0={}&\sum_{i=1}^3\bigg[-\frac{2k_i}{e_1^2h_i^2}\,\Phi_i
+\frac{2k_i}{h_j^2h_k^2}\ln x_j\ln x_k\bigg],
\label{eq:C0}\\
\Phi_i={}&\Li_2(x_j)+\Li_2(x_k)-\frac{\pi^2}{6}\nonumber\\[-2pt]
&+\ln x_j\ln y_j+\ln x_k\ln y_k
-\ln x_j\ln x_k,\nonumber
\end{align}
where $\{i,j,k\}=\{1,2,3\}$. The formula contains only dilogarithms
and logarithms, with no remaining integrations. We cross-checked it against
the known cut representation and direct in--in integration, including the
normalization; for example, $\mathcal C_0(1,1,1)=0.4876746214\ldots$.

With $\delta=\tfrac12-\nu$, equivalently $m^2/H^2=2+\delta-\delta^2$, we expand
$\mathcal C=\mathcal C_0+\mathcal C_1\delta+\mathcal C_2\delta^2+\mathcal O(\delta^3)$. The first new closed correction is
\begin{equation}
\mathcal C_1=\sum_{i=1}^3\frac{F_i}{e_1h_jh_k}\,,
\label{eq:C1}
\end{equation}
where $\{i,j,k\}$ is a permutation of $\{1,2,3\}$ and $F_i$, symmetric in $j\leftrightarrow k$, is a pure function of
uniform weight three in classical polylogarithms \cite{Goncharov:2001iea},
\begin{align}
F_i={}&\zeta_3-\Li_3\Big(\frac{y_i}{x_jx_k}\Big)-\Li_3\Big(\frac{y_jy_k}{x_jx_k}\Big)
+\Li_2\Big(\frac{y_i}{x_jx_k}\Big)L_i\nonumber\\
&+2\Li_2(x_i)\ln(x_ix_jx_k)-2\Li_2(y_i)\ln\frac{y_ix_jx_k}{x_i}\nonumber\\
&-4\Li_2\Big(\!-\frac{y_i}{x_i}\Big)\ln(x_jx_k)
+\!\!\sum_{(a,b)}\!\Big[2\Li_3\Big(\frac{y_i}{x_a}\Big)\nonumber\\
&-2\Li_3(x_a)+\Li_3\Big(\frac{y_b}{x_ix_a}\Big)-\Li_3\Big(\frac{y_iy_a}{x_ix_a}\Big)\nonumber\\
&-\Li_2\Big(\frac{y_iy_a}{x_ix_a}\Big)L_i-2\Li_2(y_a)\ln\frac{y_i}{x_b}\nonumber\\
&+2\Li_2\Big(\frac{y_a}{x_i}\Big)\ln\frac{y_i}{x_i^2x_a}+2\Li_2\Big(\frac{y_a}{x_ix_b}\Big)\ln x_i\Big]\,,
\label{eq:F}
\end{align}
with $L_i=\ln[y_i/(x_i^2x_jx_k)]$ and the sum over $(a,b)=(j,k),(k,j)$. All arguments of the trilogarithms lie in
$(0,1)$, so Eq.~\eqref{eq:F} is real and directly evaluable for every triangle; its symbol has 41 terms in the six
letters $x_1,x_2,x_3,y_1,y_2,y_3$. The ancillary files give $F_i$ in Mathematica and Python~\anc.
Equation~\eqref{eq:C1} thus has the canonical form familiar from
Feynman integrals, pure functions multiplied by leading singularities
\cite{Cachazo:2008vp,Arkani-Hamed:2010pyv,Henn:2013pwa}, \new{as at conformal coupling~\cite{Henn:2026lfz},
now away from it}.

\begin{figure}[t]
\centering
\includegraphics[width=\columnwidth]{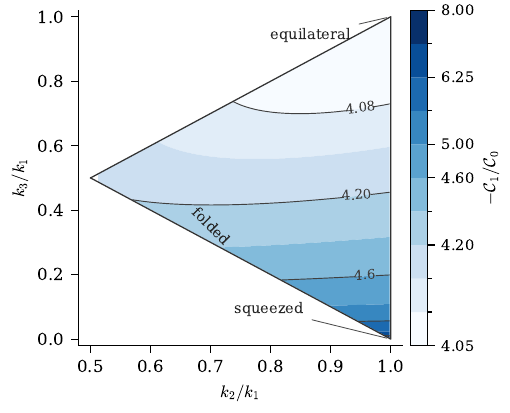}
\caption{The ratio of the first mass correction to the conformally coupled triangle, $-\mathcal C_1/\mathcal C_0$,
over the triangle shapes $k=(k_1,k_2,k_3)$ with $k_1\geq k_2\geq k_3$, from Eq.~\eqref{eq:C1} and the closed form of
$\mathcal C_0$. The ratio is nearly constant except toward the squeezed limit.}
\label{fig:shape}
\end{figure}

Near the conformally coupled mass the mass dependence is close to an overall factor: for $k_3\geq0.2\,k_1$ the ratio
$\mathcal C_1/\mathcal C_0$ lies between $-4.60$ and $-4.06$, with the smallest magnitude at the equilateral point
(Fig.~\ref{fig:shape}).

\new{The new second-order correction, valid for every physical triangle, is}
\begin{equation}
\new{\mathcal C_2=-\sum_{i}\frac{2k_i\,N_i}{(e_1h_i)^2}+\sum_{i\neq j}\frac{M_{ij}}{h_i^2h_j}\,,}
\label{eq:C2}
\end{equation}
\new{where $N_i$ and $M_{ij}$ are weight-four multiple polylogarithms built from the same six letters. Remarkably, their lengthy GPL expansions admit compact representations as noncommutative polynomials in logarithmic letters, making the uniform weight and the $k_j\leftrightarrow k_k$ symmetry of $N_i$ transparent (\EM{E}). The fully expanded standard-GPL expressions, exact rational coefficients, and numerical evaluator are provided in the ancillary files~\anc. At second order, partial fractions and residual cuts give the relevant double-pole prefactors; notably, the fitted result does not require the candidate factors containing $k_i+k_j$.} The weight thus rises
by one \new{at each of the two orders}, as at tree level \cite{Gasparotto:2024bku,Baumann:2026atn,Ahmed:2026triple}. Assigning $\delta$ the
weight $-1$, like the dimensional regulator $\epsilon$ in canonical Feynman integrals \cite{Henn:2013pwa} \new{and its analog for power-law
cosmologies~\cite{De:2023xue}}, the
expansion $\mathcal C_0+\mathcal C_1\delta+\mathcal C_2\delta^2$ has uniform weight two.

\textit{\bluehead Method}---We assemble \new{$\mathcal C$} from wavefunction coefficients with the cutting rule of
Ref.~\cite{McCune:2026vqb} (see also Refs.~\cite{Tong:2021wai,Qin:2023bjk,Stefanyszyn:2024there}): an uncut line carries the Feynman kernel
$K_\nu(yt_>)[I_\nu(yt_<)-\tfrac i\pi e^{i\pi\nu}K_\nu(yt_<)]$ and a cut line $K_\nu(yt)K_\nu(yt')$, where $t$ is the
rotated conformal time \cite{DiPietro:2021sjt} and $y$ the energy of the line.

The triangle is harder than its flat-space counterpart for three reasons. First, time is not integrated away: each vertex sits at its own time on either branch of the in-in contour, so three nested time integrals with all their orderings and cuts precede the loop integral, and since energy is not conserved at the vertices, the integrand depends on the partial energies of all subgraphs. Second, for generic mass away from $m^2=2H^2$, the propagators involve Bessel functions, so the time integrals no longer produce rational functions. Third, the loop momentum is three-dimensional, and its measure contains the square root of a Cayley--Menger determinant, as in the Baikov representation~\cite{Baikov:1996iu,Pimentel:2026kqc,Chen:2026dqp}, which already generates elliptic curves at conformal coupling~\cite{Benincasa:2024ptf,McCune:2026qfy}. Our approach removes the second difficulty order by order in the mass: at each order the Bessel-function corrections become additional auxiliary-energy integrals with rational time-integral kernels, while the first and third difficulties remain.

The key to this reduction is to expand each massive propagator around $\nu=\tfrac12$.
Differentiation with respect to $\nu$ can be represented as an auxiliary-energy \emph{dressing} of one line end~\cite{Belrhali:2026ygh,Belrhali:2026jqe,Grafe:2026dqn}. At a vertex $v$, it multiplies the conformal exponential by $e^{-2\lambda y t_v}$, shifting only the vertex energy $X_v\to X_v+2\lambda y$. The dressing has weight $1/(1+\lambda)$ for a $K_\nu$ end and $1/(\lambda-1-i0)$ for the earlier end of a Feynman line~\EM{A}. Thus the Bessel-function dependence is traded for an auxiliary-energy integral, while the ordered time integrations still yield rational energy denominators. Summing all orderings and cuts gives $\mathcal C_1$ as 72 integrals over loop energies and one dressing variable, each with four linear denominators.

The first-order dressings reduce to an exact two-fold integral; related differential-equation and alphabet-based approaches to conformal cosmological correlators provide useful context \cite{Sachs:2026dressed,Paranjape:2026cluster}. We reconstruct Eq.~\eqref{eq:C1} from a symmetric ansatz of maximal-cut prefactors and integrable symbols on the conformal alphabet; lattice reduction fixes its 114 rational parameters from 17 generic shapes at 45 digits~\EM{C} \cite{Lenstra:1982lll,Laporta:2017okg}. Independent points, alternative function representations and folded-edge integrations validate the result~\EM{D}. For Eq.~\eqref{eq:C2}, single second-order insertions and pairs of first-order dressings are both required. An independently reconstructed folded-edge function and exact regularity conditions reduce the generic weight-four ansatz to 49 unknowns. Disjoint reconstructions recover the same coefficients; six previously unseen shapes, selected and predicted before integration, agree with full-loop evaluations at about $10^{-20}$ and independent three-dimensional checks at their numerical accuracy~\EM{D}. These predictions use no data that fixed the coefficients of the reconstructed expression.

\textit{\bluehead Maximal cuts and elliptic curves}---
The compact structure of Eq.~\eqref{eq:C1} has three
noteworthy features.
First, four-fold maximal cuts constrain its rational
prefactors. For every one of the 72 dressed integrands,
setting the four energy denominators to zero gives
$B=-\tfrac12(h_a h_b h_c)^2$, with distinct
$a,b,c\in\{0,1,2,3\}$ and
$h_0=e_1=h_1+h_2+h_3$~\EM{B}.
Defining $T_{abc}=1/(h_ah_bh_c)$, the four residue
structures obey
$T_{123}=T_{012}+T_{013}+T_{023}$.
Thus only three are independent, and precisely these
prefactors survive the reconstruction~\EM{C}.
Unlike the squared folded denominators appearing in
the conformal result $\mathcal C_0$, the maximal-cut
prefactors of $\mathcal C_1$ contain only simple
folded poles.

Second, $F_i$ uses the same six-letter alphabet as
$\mathcal C_0$, with only $x_i$ appearing as first
entries of its symbol~\cite{Goncharov:2010jf}.
This is consistent with the absence of additional
physical-sheet branch points away from soft momenta
$k_i=0$. The apparent folded poles cancel in the
complete expression: $F_i$ vanishes linearly as
$y_j$ or $y_k\to0$, compensating the denominators
of Eq.~\eqref{eq:C1}~\EM{E}.
The resulting folded regularity is consistent with
the Bunch--Davies state~\cite{Arkani-Hamed:2018kmz}.

Third, intermediate loop-energy integrations encounter
elliptic geometry: on generic dressed-denominator
planes, the square root of $B$ contains an irreducible
quartic~\EM{B}. Nevertheless, the complete first
mass correction is polylogarithmic. At conformal
coupling, elliptic periods were shown to cancel in
the corresponding flat-space wavefunction
coefficient~\cite{McCune:2026qfy}.
Here the complete first mass correction to the
physical in--in correlator is likewise
polylogarithmic, despite the elliptic intermediate
geometry.

\textit{\bluehead Squeezed and folded limits}---Toward the squeezed limit $q\equiv k_3\to0$ at $k_1=k_2\equiv K$, $K^3\mathcal C_0\simeq\tfrac34\ln(K/q)$ and
$K^3\mathcal C_1\simeq-\tfrac34\ln^2(K/q)$, so that the ratio grows as $\ln(k_3/k_1)$. These leading logarithms
are those of $K^3\mathcal C\supset\tfrac3{8\delta}[1-(q/K)^{2\delta}]$, i.e., of the power $(k_3/k_1)^{1-2\nu}$, which carries twice
the mass dependence of a single $\sigma$ line: the soft leg probes the two-particle state of $\sigma$ \cite{Arkani-Hamed:2015bza,Qin:2023bjk}, whose scaling
for heavy fields gives the oscillation with frequency $2\tilde\nu$\new{, $\tilde\nu=\sqrt{m^2/H^2-9/4}$,} found in Ref.~\cite{Qin:2026yuo}. The closed form
thus reproduces the expected two-particle scaling of the loop. For equal masses and derivative vertices, the coefficient of $(k_3/k_1)^{2\delta}$ from the Mellin--Barnes pair contribution \cite{McCune:2026vqb,McCune:2026oneloop} agrees with ours. Expanding it in $\delta$ reproduces the leading and subleading squeezed logarithms of $\mathcal C_0$, $\mathcal C_1$ and $\mathcal C_2$. This agreement checks the soft sector, while Eqs.~\eqref{eq:C1} and~\eqref{eq:C2} provide the complete closed momentum dependence. \new{On a folded edge, $k_3=k_1+k_2$ with $r=k_1/k_3$, Eq.~\eqref{eq:C2}
becomes $-\tfrac12\mathcal C_1+(R_3+R_4)/[k_3^3r^2(1-r)^2]$, with classical polylogarithms $R_3$ and $R_4$ of $r$ of
weight three and four~\EM{D}, and for $k_1\to0$ it gives $\mathcal C_2\simeq\tfrac12\ln^3(k_3/k_1)/k_3^3$, the next
term of the power law above.}

\textit{Light fields}---The strongest infrared enhancement occurs near the massless point; unlike the general-mass representations and soft-limit results of Refs.~\cite{Qin:2026yuo,Borinsky:2026roj,McCune:2026vqb}, we also determine the first two coefficients of the full light-mass Laurent expansion. With
$d'=\tfrac32-\nu\simeq m^2/3H^2$, the triangle has a pole of fourth order. Each vertex contributes a late-time pole
and the soft region of one loop line a further one. At leading power the line functions factorize in time, all terms
of the cutting rule become the same product, and their coefficients sum to $-8\sin^3(\pi\nu)/\pi^3$. The next order
requires one vertex beyond leading power and the infrared-finite part of the three-line integral of frozen lines\new{~\cite{Wang:2025qfh}},
both of which we obtain in closed form~\EM{F}. With $p_3=\sum_ik_i^3$, $e_2=\sum_{i<j}k_ik_j$ and
$e_3=k_1k_2k_3$,
\begin{equation}
\mathcal C=\frac{p_3}{4e_3^2d'^4}-\frac{\Omega}{2e_3^2d'^3}+\mathcal O(d'^{-2})\,,
\label{eq:light}
\end{equation}
with $\Omega=\sum_ik_i^3\ln(e_1^2k_i/e_3)+2(1-\ln2)\,p_3-2(e_1e_2-4e_3)$.
The leading term is the local shape \cite{Komatsu:2001rj}\new{, in agreement with Ref.~\cite{An:2018tcq}}, and with $d'\simeq m^2/3H^2$ it is enhanced by $(81/4)(H/m)^8$. The leading singularity and part of the subleading logarithms agree with Ref.~\cite{An:2018tcq}; our expression determines the full subleading momentum dependence. Light fields
enhance the bubble as $(H/m)^4$ \cite{Chen:2016uwp,Chakraborty:2025myb}, a consequence of their late-time behavior
\cite{Gorbenko:2019rza,Cohen:2020php}; the triangle has four such factors\new{~\cite{An:2018tcq}}. The numerical Laurent series (Table~\ref{tab:coef}) agrees with the leading coefficient to ten digits and with
the second to $2\times10^{-6}$ at $k=(1,1.2,0.9)$, where it has its highest resolution, and to $3\times10^{-4}$ in
absolute value at the other two \new{momentum configurations}.

\begin{table}[t]
\caption{Coefficients of the two expansions at three \protect\new{momentum configurations}. Light side: $\mathcal C=\sum_nc_n\,d'^{-n}$; $c_4$ and
$c_3$ from Eq.~\eqref{eq:light}, $c_2$ and $c_1$ from the numerical Laurent series. Conformally coupled side:
$\mathcal C=\sum_n\mathcal C_n\delta^n$; \protect\new{$\mathcal C_0$, $\mathcal C_1$ and $\mathcal C_2$ from the closed forms.}}
\label{tab:coef}
\begin{ruledtabular}
\begin{tabular}{lrrr}
$k$ & $(1,1.2,0.9)$ & $(1,1,1)$ & $(1,0.8,0.6)$\\
\hline
$c_4$ & $0.7409550754$ & $0.75$ & $1.875$\\
$c_3$ & $0.4971347276$ & $0.7836046757$ & $0.1175907079$\\
$c_2$ & $-4.947313$ & $-5.90525$ & $-8.96032$\\
$c_1$ & $13.26777$ & $15.2850$ & $26.4721$\\
\hline
$\mathcal C_0$ & $0.4450454614$ & $0.4876746214$ & $0.9723862552$\\
$\mathcal C_1$ & $-1.8121950537$ & $-1.9777568635$ & $-3.9918905898$\\
$\mathcal C_2$ & $5.1649668877$ & $5.6190888070$ & $11.450212754$\\
\end{tabular}
\end{ruledtabular}
\end{table}

\begin{figure}[t]
\centering
\includegraphics[width=\columnwidth]{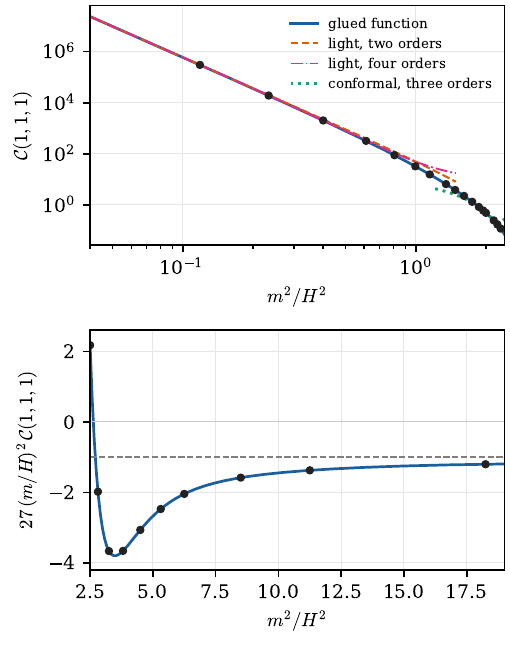}
\caption{The triangle at the equilateral point as a function of the mass. Points: direct evaluations; solid line:
Eq.~\eqref{eq:glue}. Top: light to conformally coupled masses, with the light expansion through two orders,
Eq.~\eqref{eq:light}, and through four orders (Table~\ref{tab:coef}), and the expansion around the conformally coupled
mass through second order in $m^2/H^2-2$. Bottom: $27(m/H)^2\,\mathcal C$, which tends to $-1$ for heavy fields
(dashed); the triangle changes sign at $m^2\simeq2.626H^2$.}
\label{fig:mass}
\end{figure}

\textit{From light to heavy fields}---The correlator depends on the mass only through $m^2$, so $\mathcal C$ is even
in $\nu$: with the Feynman kernel above this follows from $I_{-\nu}=I_\nu+(2/\pi)\sin(\pi\nu)K_\nu$. As a function of
$m^2$ it is analytic for $m^2>0$, and its singularity at $m^2=0$ appears as the poles at $\nu=\pm\tfrac32$.
Subtracting their principal parts $P$,
\begin{equation}
\mathcal C(\nu)=P(\nu)+P(-\nu)+f(\nu^2)\,,
\label{eq:glue}
\end{equation}
leaves a function $f$ that is smooth for all physical masses, $\nu^2<\tfrac94$ including imaginary $\nu$. In
$P(\nu)=\sum_{n=1}^4c_n(\tfrac32-\nu)^{-n}$ the coefficients $c_4$ and $c_3$ are those of Eq.~\eqref{eq:light}, and
$c_2$ and $c_1$ are read off from the numerical Laurent series. We represent $f$ by an interpolation in a basis of
twelve Chebyshev polynomials in $(m^2/H^2+3)^{-1}$, fitted to values of $\mathcal C$ on circles around $\nu=\tfrac12$
and $\nu=\tfrac32$, on the real axis for $0.55\leq\nu\leq1.46$, and for $\nu^2<\tfrac14$ up to $m^2=18.25H^2$,
together with $\mathcal C_0$ and $\mathcal C_1$; it is not used beyond this range. Values on the real axis left out of
the fit are reproduced to $2\times10^{-7}$ at $k=(1,1.2,0.9)$ and to $5\times10^{-6}$ at the equilateral point;
values for heavier fields left out of the fit are reproduced there to $10^{-4}$ up to $m^2=6.25H^2$ and to
$5\times10^{-4}$ at $m^2=8.5H^2$ (Fig.~\ref{fig:mass}).

For heavy fields, integrating out $\sigma$ at one loop produces the operator $\delta\dot\phi^3$, with a coefficient
fixed by the third derivative of the one-loop effective potential \cite{Coleman:1973jx}, $1/(32\pi^2m^2)$; this gives
$\mathcal C\to-(H/m)^2/e_1^3$~\EM{G}, as for the one-loop matching of a non-derivative coupling
\cite{Cespedes:2025ple}, and consistent with the $1/m^2$ scaling of the effective coupling that controls the soft
signal \cite{Qin:2026yuo}. The direct evaluations approach this limit\new{ (Fig.~\ref{fig:mass}, bottom)}, and the ratio to the limit agrees between $k=(1,1,1)$ and
$(1,1.2,0.9)$ to $2\times10^{-3}$ for $m^2\geq3.25H^2$. Since the triangle is positive at conformal coupling and
negative for heavy fields, it must change sign in between: the direct evaluations bracket the zero between
$m^2=2.50H^2$ and $2.81H^2$, and Eq.~\eqref{eq:glue} places it at $m^2\simeq2.626H^2$ for the equilateral triangle.

\textit{\bluehead Checks}---At the integrand level the second-order mass expansion agrees with direct Bessel-function differentiation~\EM{A}. The reconstructed $\mathcal C_2$ passes exact symmetry, scaling and folded-edge constraints; its frozen predictions agree with six blinded complete integrations at the $10^{-20}$ level and ten independent three-dimensional integrations within their numerical precision~\EM{D}. We also reproduce the conformal triangle in the equivalent flat-space and de Sitter formulations \cite{Pimentel:2026kqc,Henn:2026lfz} and the general-mass benchmarks of Refs.~\cite{Borinsky:2026roj,McCune:2026vqb}.

\textit{Size of the signal}---With $g=cH^2/\Lambda^2$, $P_{\rm sr}=|\dot\phi_0|/H^2=(2\pi A_s^{1/2})^{-1}$ and the
shape normalized at the equilateral point, the triangle has the amplitude \cite{Borinsky:2026roj}
\begin{equation}
f_T=-\frac{5\,g^3\,\mathcal C(1,1,1;\nu)}{72\,\pi^6A_s^2}\simeq-1.6\times10^{13}\,g^3\,\mathcal C(1,1,1;\nu)\,.
\label{eq:fT}
\end{equation}
For light fields Eq.~\eqref{eq:light}
gives $f_T\simeq-1.0\times10^{15}\,g^3(H/m)^8$ with the local shape. This expansion holds as long as the duration of
inflation does not cut off the late-time and soft regions, that is for $m^2/H^2\gg3/(2N)$ with $N$ the number of
$e$-folds. The operator that produces the vertex also shifts the mass by $|\delta m^2|=2|g|P_{\rm sr}^2H^2$. Relative
to the bubble, which enters at order $g^2P_{\rm sr}^2$ and requires renormalization \cite{Borinsky:2026roj}, the
triangle is therefore weighted by $\new{|g|}P_{\rm sr}^2=|\delta m^2|/\new{(2H^2)}$\new{~\cite{Kumar:2019ebj}}, and it competes with the bubble whenever the
mass shift is of order $H^2$. Without tuning\new{~\cite{An:2018tcq,Kumar:2019ebj,Wang:2019gbi,Wang:2021qez,Borinsky:2026roj}}, $|\delta m^2|\lesssim m^2$, so that $|g|\lesssim(m/H)^2/(2P_{\rm sr}^2)$
and $|f_T|\lesssim7\times10^{-8}(H/m)^2$: the enhancement $(H/m)^8$ is largely compensated by the smaller coupling
that a light mass allows. The closed forms therefore provide precision benchmarks and expose the analytic structure of massive loops even when naturalness suppresses their observable amplitude.

\textit{Conclusions}---We have obtained the first two closed analytic mass corrections to the physical one-loop inflationary triangle at generic momenta, extending the conformally coupled answer into a massive theory. Their pure weight-three and weight-four multiple polylogarithms use the same letters as the conformal triangle, with rational prefactors constrained by loop residues. That a massive cosmological loop retains this canonical structure despite elliptic intermediate geometry is the central surprise. The explicit second-order formula and its independent numerical tests turn this observation into a benchmark for future calculations. The auxiliary-energy construction extends to higher mass orders and other loop topologies, where the persistence---or failure---of this polylogarithmic structure can now be investigated.

\textit{Acknowledgments}---We thank Sumit Banik and Andreas Rapakoulias for collaboration on related projects. The work of \new{S.M.H.} is supported by the Academic Research Sabbatical Program (ARSP)
of the University of Regensburg.

\textit{\bluehead Data availability}---\new{The ancillary files give $\mathcal C_1$ and $\mathcal C_2$ for general triangles and on the folded edge in Mathematica and Python~\anc.}

\bibliography{refs}

\clearpage
\onecolumngrid
\section*{End Matter}
\twocolumngrid

\emhead{A}{Exact time integrals and the dressing}At the conformally coupled mass the lines
are elementary. A cut and an uncut line between the times $t$ and $t'$ read
\begin{equation}
G_{\rm cut}=\frac{\pi\,e^{-y(t+t')}}{2y\sqrt{tt'}}\,,\qquad
G_{\rm uncut}=\frac{e^{-y|t-t'|}}{2y\sqrt{tt'}}\,.
\label{eq:lines}
\end{equation}
With one inflaton leg per vertex, the measure $t e^{-k_vt}$ cancels the powers of $t$, so ordered time integrals are rational in the energies. To retain this property beyond conformal coupling, define the earlier-end function
$E_\nu(z)=I_\nu(z)-(i/\pi)e^{i\pi\nu}K_\nu(z)$, so an uncut line is proportional to $K_\nu(yt_>)E_\nu(yt_<)$.
Its two normalized order derivatives at $\nu=\tfrac12$ have the auxiliary-energy representations~\cite{NIST:DLMF}
\begin{equation}
\begin{aligned}
\left.\partial_\nu\ln K_\nu(z)\right|_{\frac12}
 &=\int_0^\infty\!\dd\lambda\,\frac{e^{-2\lambda z}}{1+\lambda},\\
\left.\partial_\nu\ln E_\nu(z)\right|_{\frac12}
 &=\int_0^\infty\!\dd\lambda\,\frac{e^{-2\lambda z}}{\lambda-1-i0}.
\end{aligned}
\label{eq:dressing}
\end{equation}
The second prescription includes the continuation of the phase in $E_\nu$. Thus a derivative dresses one line end at vertex $v$ by $e^{-2\lambda yt_v}$, shifting only that vertex energy, $X_v\to X_v+2\lambda y$. For instance, with $t_1=s$, $t_2=s+u$, $t_3=s+u+v$, a dressing at vertex 2 acts on the ordered time integral as
\begin{equation}
\frac{1}{R_0R_1R_2}\ \longrightarrow\
\frac{1}{(R_0+2\lambda y)(R_1+2\lambda y)R_2}.
\label{eq:dressedtime}
\end{equation}
Hence the time integrations remain rational at each mass order. The construction extends through the Legendre representation~\cite{Erdelyi:1954}
\begin{equation}
K_\nu(z)=\sqrt{\frac{\pi z}{2}}\int_1^\infty\!\dd x\,e^{-zx}P_{\nu-\frac12}(x),
\end{equation}
underlying the massive dressing method~\cite{Belrhali:2026ygh,Belrhali:2026jqe,Grafe:2026dqn}. In the Taylor expansion in $\nu-\tfrac12$, the second-order single-end weights are $\ln(1+\lambda)/\lambda$ for $K_\nu$ and $\ln(1-\lambda+i0)/\lambda$ for $E_\nu$. Together with products of first-order dressings, they produce the complete $\mathcal C_2$ integrand, checked against direct Bessel-function integration at fixed loop momentum.

\emhead{B}{The integrand and its maximal cuts}We integrate over the energies $D_i$ of the
three single-vertex tubes. In these variables the loop measure is $\dd^3D\,B^{-1/2}$, where $B$ is the Cayley--Menger
determinant of the loop momentum and the three line centres, quadratic in each $D_i$. A dressed line $L$ adds the
variable $\omega=2\lambda y_L$, and each of the 72 terms of $\mathcal C_1$ takes the form
\begin{equation}
\int\frac{\dd^3D}{\sqrt{B}}\int_0^\infty\!\dd\omega\,\prod_{n=1}^4\frac{1}{f_n}\,,
\label{eq:term}
\end{equation}
with four linear forms $f_n$ drawn from $D_i$, $D_i+h_i$, $D_i+\omega$, $D_i+h_i+\omega$, $e_1+\omega$ and
$\omega\pm2y_L$. At the common zero of the four forms the Jacobian is $\pm1$, and the maximal cut is
\begin{equation}
B=-\tfrac12\,(h_ah_bh_c)^2\,,
\label{eq:maxcut}
\end{equation}
with $\{a,b,c\}=\{0,1,2\}$, $\{0,1,3\}$ or $\{0,2,3\}$ for 20 terms each and $\{1,2,3\}$ for the remaining 12. No
elliptic curve appears on the cuts: on the 288 lines where three forms of a term vanish, $B$ is quadratic (180 lines)
or the square of a quadratic (108 lines). Elliptic curves appear only in the integrations. On the plane $y_L=0$, $B$
restricted to the two remaining loop energies is an irreducible quartic, whose quartic part defines the curves
\begin{equation}
w^2=k_a^2\tau^4-(k_a^2+k_b^2-k_c^2)\,\tau^2+k_b^2
\end{equation}
of the conformally coupled triangle \cite{McCune:2026qfy}, with $\tau$ the ratio of the two loop energies. On most
planes where a dressed denominator vanishes, $B$ is an irreducible quartic as well.

\emhead{C}{Two-fold representation and the fit}In each term of Eq.~\eqref{eq:term} the
dressed energy is the only quantity that changes along the redundant sum of loop energies. That integral and the
dressing integral therefore close into logarithms of algebraic functions, and two integrals over sums of loop
energies remain. We evaluate them with ball arithmetic \cite{Johansson:2017arb} and double-exponential quadrature
\cite{Takahasi:1974} to 50 or 100 digits, in about half a minute per configuration at 50 digits on 64 cores.

The ansatz for $\mathcal C_1$ multiplies the prefactors of $\mathcal C_0$ and the maximal cuts by all integrable
symbols of weight three in the letters $x_i,y_i$ with first entries $x_i$. We realize them as two-dimensional harmonic
polylogarithms \cite{Remiddi:1999ew,Gehrmann:2000zt}, evaluated with GiNaC \cite{Bauer:2000cp,Vollinga:2004sn}, and
multiply functions of weight $w<3$ by constants of weight $3-w$ built from $\ln2$, $\pi^2$ and $\zeta_3$.
Symmetrized over permutations of the momenta, the ansatz has 114 integer coefficients. Lattice reduction on values at
17 generic configurations returns a single vector; two disjoint sets of 18 configurations at 40 and 30 digits return
the same vector, and a planted solution of the same size is recovered exactly. Only the maximal cuts receive nonzero
coefficients.

\emhead{D}{Folded values and independent validation}At $h_i=0$ the line centres become
collinear. Prolate spheroidal coordinates give an independent two-fold
representation at folded momenta, yielding high-precision values and exact
relations in logarithms, zeta values and polylogarithms. For instance,
\begin{align}
96\,\mathcal C_1(2,1,1)={}&12-36\ln2-60\ln^22-2\pi^2\nonumber\\
&-4\ln^32-2\pi^2\ln2-9\zeta_3.
\label{eq:folded}
\end{align}
The ancillary files give the exact values of $\mathcal C_2(2,1,1)$ and $\mathcal C_2(2/3,1/3,1)$~\anc. The second-order folded \emph{function}
fixes the edge constraints, leaving 49 coefficients. Ten complete
28-digit loop values fix these coefficients; disjoint reconstructions,
nine further values, ten independent three-dimensional evaluations and
six blinded full-loop tests confirm the expression.

\emhead{E}{Closed weight-four numerators and their symmetry}Equation~\eqref{eq:F} gives the
symmetric $F(y_j,y_k)$; its two equivalent forms agree to 60 digits. Each $F_i$ vanishes
linearly as $y_j$ or $y_k\to0$, so Eq.~\eqref{eq:C1} is regular at folded kinematics.
The exact GPL coefficients factor into compact noncommutative word
polynomials, using the standard iterated-integral algebra
\cite{Chen:1977ipa,Duhr:2011zq}. Define the map $\mathcal I$ by
\begin{equation}
\dd\mathcal I[z_1\cdots z_n]=\mathcal I[z_1\cdots z_{n-1}]\,\dd\ln z_n,
\quad\mathcal I[\varnothing]=1,
\label{eq:iterletters}
\end{equation}
with the straight path in the ratios $(k_i/k_k,k_j/k_k,1)$ from
$(2,1,1)$ to their physical values, regularized at the folded base point. The map is linear; products inside $\mathcal I$ denote \emph{ordered words}.
Along the straight path $t\in[0,1]$, let
$\dd\ln L_s=\sum_\alpha m_{s\alpha}\,\dd t/(t-r_{s\alpha})$.
Then, in standard GPL notation~\cite{Goncharov:2001iea},
\begin{equation}
\mathcal I[L_1\cdots L_n]=\sum_{\boldsymbol\alpha}
\Big(\prod_{s=1}^n m_{s\alpha_s}\Big)
G(r_{n\alpha_n},\ldots,r_{1\alpha_1};1).
\label{eq:wordGPL}
\end{equation}
The order reverses because $L_1$ is integrated first. Set
\begin{equation}
\begin{gathered}
A=\frac{k_jk_k}{e_1^2},\quad B=\frac{h_i}{e_1},\quad
C=\frac{h_jh_k}{k_jk_k},\\
D=\frac{k_j}{k_k},\qquad E=\frac{k_jh_j}{k_kh_k}.
\end{gathered}
\label{eq:NL}
\end{equation}
Then the first numerator of Eq.~\eqref{eq:C2} is
\begin{align}
N_i=\mathcal I\big[&P_4+P_3\ln2+P_2(\ln^22+\tfrac{\pi^2}{6})
+\tfrac{\pi^2}{3}C^2\nonumber\\
&-C(\tfrac{13}{4}\zeta_3+\tfrac{\pi^2}{3}\ln2+\tfrac23\ln^32)\big],
\label{eq:Ni}
\end{align}
where
\begin{align}
P_2={}&AC-\tfrac52C^2+DE-\tfrac12E^2,\nonumber\\
P_3={}&(C-A)P_2+2C(A-B+C)C\nonumber\\
&+\tfrac12(3D-E)(CE+EC),\nonumber\\
P_4={}&-\tfrac12AP_3+\tfrac14D(C-A)(CE+EC)\nonumber\\
&+\tfrac12D(3D-E)P_2
+D[4DC-E(A-B+C)]C.
\label{eq:P1234}
\end{align}
Under $k_j\leftrightarrow k_k$, $A,B,C$ are unchanged while $D,E$
are inverted. Every word has an even total number of $D,E$ factors,
manifesting $N_i(k_i;k_j,k_k)=N_i(k_i;k_k,k_j)$.
For the second numerator introduce
\begin{equation}
\begin{gathered}
a=\frac{k_j}{h_i},\quad b=\frac{k_j}{e_1},\quad
c=\frac{e_1}{k_i},\\
d=\frac{h_j}{k_i},\qquad e=\frac{e_1h_k}{k_ik_j}.
\end{gathered}
\label{eq:ML}
\end{equation}
Then
\begin{align}
M_{ij}=\mathcal I\big[&Q_4+Q_3\ln2
+Q_2(\ln^22+\tfrac{\pi^2}{6})\nonumber\\
&-\tfrac{\zeta_3}{4}(2c+15e)
+(c-e)(\tfrac{\pi^2}{6}\ln2+\tfrac13\ln^32)\big],
\label{eq:Mij}
\end{align}
with
\begin{align}
Q_2={}&-bc+de-e^2,\nonumber\\
Q_3={}&2(d-2b)Q_2+4(d-b)e^2+2c(bc-ae),\nonumber\\
Q_4={}&-bQ_3+2c[(a-2c)(ae+cb)\nonumber\\
&-ae^2-b(cb+de)].
\label{eq:Q234}
\end{align}
The potentially singular letters $B$ and $a$ never occur first, so the
base-point integrals converge. The ancillary files give the explicit $G$-function expressions for $N_i$ and $M_{ij}$~\anc.

\emhead{F}{The two light orders}Near $d'=\frac32-\nu=0$, each
vertex contributes a late-time pole and one soft loop region supplies
a fourth pole. At the massless point $K_{3/2}(z)=
\sqrt{\pi/(2z)}e^{-z}(1+1/z)$. The soft-region residue and the
infrared-finite frozen-line integral determine the first two Laurent
coefficients in Eq.~\eqref{eq:light}. In particular, the momentum-space
vertex residues have
\begin{equation}
r_3=\frac{o_l^3}{8e_3^2},\quad
r_2=r_3\!\left(4-2\ln2+2\ln e_1-\frac{2e_1}{o_l}
+\frac{2e_3}{o_l^2e_1}\right),
\end{equation}
where $o_l$ is opposite line $l$. The remaining finite radial integral
agrees with the known result of Ref.~\cite{Wang:2025qfh} in its
regulator-independent terms.

\emhead{G}{The heavy-mass limit}Since $(\partial\phi)^2\supset-2\dot\phi_0\,\delta\dot\phi$,
the inflaton fluctuation shifts the mass of $\sigma$,
\begin{equation}
m^2\to m^2+J\,,\qquad J=\frac{4c\,\dot\phi_0\,\delta\dot\phi}{\Lambda^2}\,.
\end{equation}
For $m\gg H$ we integrate out $\sigma$ with the one-loop effective potential $V=M^4[\ln(M^2/\mu^2)-\tfrac32]/(64\pi^2)$,
which gives the local operator
\begin{equation}
\mathcal L_{\rm eff}\supset-\frac{V'''}{6}\,J^3=\lambda\,\delta\dot\phi^3\,,\qquad
\lambda=-\frac{c^3\dot\phi_0^3}{3\pi^2\Lambda^6m^2}\,,
\end{equation}
since $V'''=1/(32\pi^2m^2)$; the coefficient is independent of the renormalization scale. The in-in integral of this
contact operator gives $\langle\delta\phi_{\vec k_1}\delta\phi_{\vec k_2}\delta\phi_{\vec k_3}\rangle'=3\lambda
H^5/(k_1k_2k_3e_1^3)$, and comparison with Eq.~\eqref{eq:def} yields
\begin{equation}
\mathcal C\to-\Big(\frac Hm\Big)^2\frac{1}{e_1^3}\,,
\end{equation}
with corrections of relative order $H^2/m^2$.

\end{document}